\documentclass[twocolumn,prl,groupedaddress,showpacs,amsfonts,amssymb,amsmath,letterpaper]{revtex4-1}
\usepackage{amsmath}
\usepackage{graphicx}
\usepackage{amssymb}
\usepackage{color}

\makeatletter
\begin{document}
\title{
Waveguide-QED-Based Photonic Quantum Computation }

\author{Huaixiu Zheng}
\email{hz33@duke.edu}
\author{Daniel J. Gauthier}
\author{Harold U. Baranger}
\email{baranger@phy.duke.edu} \affiliation{\textit{Department of
Physics, Duke University, P.\,O.\,Box 90305, Durham, North Carolina
27708, USA}}

\date{August 21, 2013}

\begin{abstract}
We propose a new scheme for quantum computation using flying qubits\textemdash propagating photons in a one-dimensional waveguide interacting with matter qubits. Photon-photon interactions are mediated by the coupling to a four-level system, based on which photon-photon $\pi$-phase gates (Controlled-NOT) can be implemented for universal quantum computation. 
We show that high gate fidelity is possible given recent dramatic experimental progress in superconducting circuits and photonic-crystal waveguides. The proposed system can be an important building block for future on-chip quantum networks.
\end{abstract}


\maketitle Quantum computers hold great promise for outperforming
any classical computer in solving certain problems such as integer
factorization \cite{
Nielsen00}, as well as in efficiently simulating quantum many-body systems 
\cite{
BulutaSci09, CiracNatPhys12}. While quantum
computation schemes often encode information in stationary qubits
such as atoms, trapped ions, quantum dots and superconducting qubits 
\cite{LaddNat10}, flying qubits\textemdash photons\textemdash have several
appealing features as carriers of quantum information \cite{LaddNat10,PetrosyanJOpt05}.
Most importantly, photons have long coherence times because they rarely interact, and yet can be readily manipulated at the single photon level using linear optics. Furthermore, photonic quantum computation is potentially scalable \cite{PetrosyanJOpt05} in view of the recent controlled generation of single-photon pulses \cite{SantoriNat02, ChouPRL04, ClaudonNatPhoton10, LauchtPRX12, EichlerPRL11} and demonstration of stable quantum memories \cite{HedgesNat10,ReimPRL11}. However, weak photon-photon interaction makes it very challenging to realize the two-qubit gates necessary for universal computation between single-photons \cite{PetrosyanJOpt05,ShapiroPRA06,FanPRL13}. Several schemes have been proposed to circumvent this difficulty. The linear optics scheme \cite{KnillNat01} uses quantum interference between qubit photons and auxiliary photons to generate an effective nonlinear interaction between qubit photons. Other approaches include employing trapped atoms in a cavity \cite{DuanPRL04, KoshinoPRA10} or Rydberg atoms \cite{GorshkovPRL11,PeyronelNat12} to realize two-qubit gates.

In this work, we propose an alternative scheme for photonic quantum computation: using strong coupling between local emitters and photons in a one-dimensional
(1D) waveguide. Because of recent tremendous experimental progress
\cite{AkimovNat07, BajcsyPRL09, BabinecNatNanotech10,ClaudonNatPhoton10, AstafievSci10,AstafievPRL10,BleusePRL11,HoiPRL11,LauchtPRX12,EichlerPRL11}, 1D waveguide
systems are becoming promising candidates for quantum information processing. 
A variety of capabilities has been proposed \cite{ChangNatPhy07,ShenPRL07,ZhouPRL08,LongoPRL10,WitthautNJP10,ZhengPRA10,RoyPRL11,KolchinPRL11,ZhengPRL11,*ZhengPRA12}, particularly at the single photon level, yet protocols compatible with current waveguide setups for some important tasks, notably two-qubit gates, have rarely been investigated \cite{CiccarelloPRA12}. 
In our proposal, we construct photonic two-qubit gates solely based on scattering in a waveguide system that is accessible in current experiments. 
The photonic qubits are initialized by and returned to quantum memories \cite{HedgesNat10,ReimPRL11} in order to realize long-term storage.
Compared with the cavity approach, our setup is simplified and avoids the complexity of stabilizing the resonance between the cavity modes and the atom. The gate has a wide bandwidth, and its operation time is determined solely by the coupling strength. Combining the simplicity of the system and the scalability of photons, our waveguide-QED-based scheme opens a new avenue towards scalable quantum computation and distributed quantum networks \cite{KimbleNat08} in a cavity-free setting.

\begin{figure}[t!]
 \centering
 \includegraphics[width=0.5\textwidth]{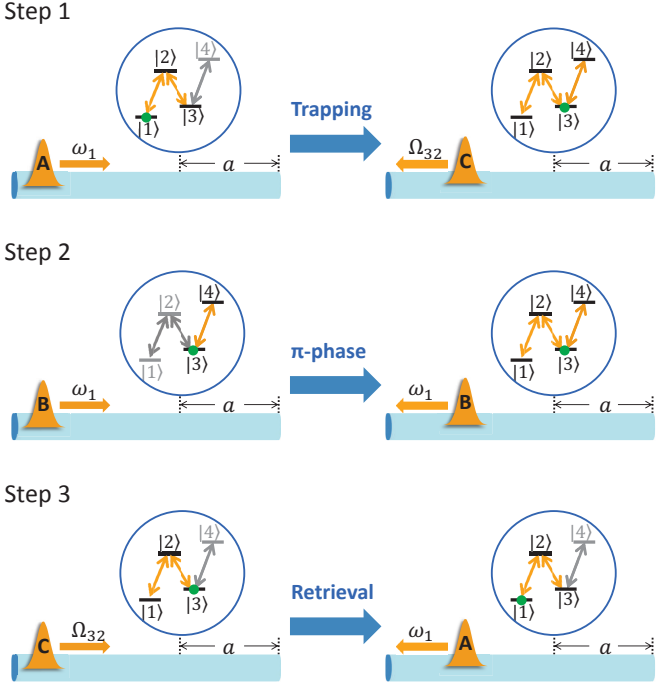}
\caption{Gate operation: 1) trapping; 2) $\pi$-phase; 3)
retrieval. The gate sequences here illustrate the case of both photons $A$ and
$B$ being in state $|\omega_1\rangle$. For this case, step 4) does not cause any change and hence is not shown. The left and right sides show
the initial and final states, respectively. Inactive transitions in
each step are gray-colored.}
 \label{fig:Scheme2}
 \end{figure}

The photonic qubits are encoded in the frequency degree of freedom $|\omega_0\rangle$ and $|\omega_1\rangle$ for simplicity; a straightforward generalization of our scheme is applicable to polarization-encoded qubits \footnote{See Supplemental Material for the generalization to polarization-qubit, derivations of results in Eqs.\,(2)-(8), a $M$-level scheme for a qubit-photon quantum memeory, and an alternative 3LS scheme for $\pi$-phase gates.}. 
Single photons can be generated from the emission of quantum dots \cite{ClaudonNatPhoton10, BleusePRL11, LauchtPRX12} or using circuit-QED systems \cite{EichlerPRL11}, and single-qubit rotations can be realized using a Mach-Zehnder interferometer \cite{MadsenPRL06,BozyigitJPhysConf11}. 
Hence, we focus on two-qubit gates and, in particular, a $\pi$-phase (Controlled-{\small NOT}) gate. 
We consider a semi-infinite 1D waveguide side-coupled to a four-level system that is located a distance $a$ from the end (Fig.\,\ref{fig:Scheme2}). 
Such a setup can be realized in a variety of experimental systems using 
superconducting transmission lines \cite{AstafievSci10, AstafievPRL10,HoiPRL12,EichlerPRL11}, 
diamond nanowires coupled to NV centers \cite{BabinecNatNanotech10}, photonic-crystal waveguides coupled to quantum dots \cite{LauchtPRX12}, hollow fibers with trapped cold atoms \cite{BajcsyPRL09}, or plasmonic nanowires \cite{AkimovNat07}. 
We now show that a $\pi$-phase gate between two photons $A$ and $B$ can be realized by reflecting them from the end of the waveguide.

The Hamiltonian of the system (Fig.\,\ref{fig:Scheme2}) is given by
\begin{eqnarray}
\lefteqn{H =H_{\rm wg}+\sum^4_{i=2}\hbar(\Omega_i-i\Gamma^{\prime}_i/2)\sigma_{ii}
}
\nonumber \\
&&+\!\!\sum_{\alpha=R,L}\int \!dx\, \hbar
V\,\delta(x)[a^{\dagger}_{\alpha}(x)(\sigma_{12}+\sigma_{32}+\sigma_{34})+\textrm{h.c.}],\;\;\nonumber\\
&&H_{\rm wg}=\int dx\frac{\hbar
c}{i}\Big[a_{R}^{\dagger}(x)\frac{d}{dx}a_{R}(x)-a_{L}^{\dagger}(x)\frac{d}{dx}a_{L}(x)\Big],
\end{eqnarray}
where $a_{R,L}(x)$ are the propagation modes along the $x$ axis of the waveguide, $\sigma_{ij}\equiv |i\rangle\langle j|$, and the energy of ground state $|1\rangle$ is the energy reference. 
An imaginary term models the loss of the excited state at rate $\Gamma^{\prime}_i$. The
decay rate to the waveguide continuum is $\Gamma = 2V^2/c$,
where c is the group velocity of photons.
For our gate operation, we require that the transitions
$1\!\rightarrow\!2$ and $3\!\rightarrow\!4$ have the same frequency
$\Omega_{12}=\Omega_{34}$ (where $\Omega_{ij}\equiv\Omega_j-\Omega_i$);
in contrast, the frequency of the $3\!\rightarrow\!2$ transition should be distinctly
different, satisfying $|\Omega_{32}-\Omega_{12}|\gg
\Gamma$.
In addition, for simplicity, we assume that (i)~transitions
$1\!\rightarrow\! 2$, $3\!\rightarrow\! 2$, and $3\!\rightarrow\! 4$ have the
same coupling strength $\Gamma$ to the waveguide modes,
(ii)~state $3$ is metastable with loss rate $\Gamma^{\prime}_{3}=0$, and 
(iii)~states $2$ and $4$ have the same loss rate 
$\Gamma^{\prime}\equiv\Gamma^{\prime}_2=\Gamma^{\prime}_4$. None of these additional assumptions is essential. Here, we set $\hbar=c=1$.

The photon qubit consists of two distinct frequencies.
Frequency $\omega_1$ is chosen to be on resonance with the
transitions $1\rightarrow2$ and $3\rightarrow4$,
\textit{i.e.} $\omega_1=\Omega_{12}$. In contrast, $\omega_0$ is far off resonance from all
of the atomic transitions---an $\omega_0$ photon
does not interact with the four-level system (4LS). The 4LS is initialized in $|1\rangle$.
Here, we assume that we have quantum memories \cite{HedgesNat10,ReimPRL11} available, one for qubit photon $A$ and another for $B$, and we give an example in the Supplementary Materials \cite{Note1}.

As illustrated in Fig.\,\ref{fig:Scheme2}, a $\pi$-phase gate
between photon pulses $A$ and $B$ can be realized via the following
steps. 
(1)~Trapping: photon $A$ (of frequency $\omega_A$) is sent into
the system. If $A$ is in state $|\omega_1\rangle$, it is trapped: 
the 4LS makes the transition from $|1\rangle$ to $|3\rangle$ with photon $A$ being stored in state $|3\rangle$. 
In addition, an auxiliary photon $C$ of frequency $\Omega_{32}$ is
emitted, and the 4LS is put into $|3\rangle$. Otherwise, $A$ will come out without interacting with the 4LS. We call the output the $A^{\prime}$ photon.
(2)~$\pi$-phase: a
second qubit photon $B$ (of frequency $\omega_B$) is sent into the system; it gains a
$\pi$-phase if both $\omega_B=\omega_1$ and the 4LS
is in state $|3\rangle$. Otherwise, $B$ will either pass through without any change if $\omega_B=\omega_0$, or be trapped followed by the emission of a $C^{\prime}$-photon of frequency $\Omega_{32}$ if $\omega_B=\omega_1$ and the 4LS is in state $|1\rangle$. 
(3)~Retrieval of A: by time reversal arguments, sending in the output
photon $A^{\prime}$ retrieves photon $A$, which is further directed to and stored in quantum memory $A$.
(4)~Retrieval of B: in the case of $\omega_A=\omega_0$ and $\omega_B=\omega_1$, photon $B$ will be trapped by the 4LS in step 2---we retrieve it by simply sending in the auxiliary photon $C$. In all the other cases, photon $C$ simply passes through the 4LS without any change. Photon $B$ from either step 2 or 4 is directed to and stored in quantum memory $B$.
Therefore, only when $\omega_A=\omega_B=\omega_1$ is a $\pi$-phase generated by their
interaction with the 4LS.
We now analyze each of these steps.

\emph{Step 1\textemdash Trapping}. For an incoming single photon $A$ in mode $|\omega_A\rangle$ and initial
state $|1\rangle$ of the 4LS, the output state of the system, obtained by imposing wavefunction matching \cite{ZhengPRL11,*ZhengPRA12} and a hard-wall boundary condition at the end of the waveguide (we assume perfect reflection from the end of the waveguide) \cite{CiccarelloPRA12, Note1}, is 
\begin{equation}
|\phi^{\text{out}}_1(\omega_A)\rangle=r_{11}(\omega_A)|\omega_A\rangle\otimes |1\rangle +
r_{13}(\omega_A)|\tilde{\omega}_A\rangle\otimes |3\rangle,
\label{phi1}
\end{equation}
where 
\begin{eqnarray}
\tilde{\omega}&=&\omega-\Omega_{13}, \nonumber \\
r_{11}(\omega)&=&e^{2i\omega
a}\frac{-\Omega_{12}+\frac{i\Gamma^{\prime}}{2}+\omega-\frac{i\Gamma}{2}[e^{2i\tilde{\omega}a}-e^{-2i\omega
a}]}{\Omega_{12}-\frac{i\Gamma^{\prime}}{2}-\omega+\frac{i\Gamma}{2}[e^{2i\tilde{\omega}a}+e^{2i\omega
a}-2]}, \nonumber \\
r_{13}(\omega)&=&\frac{(i\Gamma/2)(e^{2i\omega
a}-1)[e^{2i\tilde{\omega}a}-1]}{\Omega_{12}-\frac{i\Gamma^{\prime}}{2}-\omega+\frac{i\Gamma}{2}[e^{2i\tilde{\omega}a}+e^{2i\omega
a}-2]} \;.
\end{eqnarray}
We first illustrate the operation principle for the lossless case $\Gamma^{\prime}=0$ and then later analyze the effect of loss in detail.
We assume that the key condition $2(\Omega_{12}+\Omega_{32})a=2n_1\pi$ is satisfied; in addition, we can make the trivial choice $2\omega_0a=(2n_0+1)\pi$ ($n_0,n_1$ are integers).
Then, if the incoming qubit photon-$A$ is
in mode $|\omega_0\rangle$, $r_{11}(\omega_0)=1$ and $r_{13}(\omega_0)=0$ because $\omega_0$ is far off resonance from all the transitions. Hence, it will reflect from the system without change, leaving the 4LS in $|1\rangle$.
On the other hand, if photon-$A$ is in mode $|\omega_1\rangle$, the on-resonance interaction with the $1\rightarrow2$ transition gives $r_{11}(\omega_1)=0$ and $r_{13}(\omega_1)=-1$. As a result, it will be trapped and stored in level $|3\rangle$ of the 4LS, emitting an auxiliary $C$-photon at frequency $\omega_1-\Omega_{13}=\Omega_{32}$.

\emph{Step 2\textemdash $\pi$-phase}. Now send in the second qubit, photon-$B$ in mode $|\omega_B\rangle$. 
The output state after scattering reads \cite{Note1}

\begin{eqnarray}
|\phi^{\text{out}}_2(\omega_A,\omega_B)\rangle&=&
r_{11}(\omega_A)|\omega_A\rangle \otimes |\phi^{\text{out}}_1(\omega_B)\rangle
   \nonumber \\
&+&r_{13}(\omega_A)R_{3}(\omega_B)|\tilde{\omega}_A\rangle |\omega_B\rangle\otimes|3\rangle,
\end{eqnarray}

where
\begin{equation}
 R_{3}(\omega)=\frac{-(\Omega_{12}-\frac{i\Gamma^{\prime}}{2}-\omega)e^{2i\omega a}+\frac{i\Gamma}{2}(1-e^{2i\omega a})}{\Omega_{12}-\frac{i\Gamma^{\prime}}{2}-\omega-\frac{i\Gamma}{2}(1-e^{2i\omega a})}.
\end{equation}
Here, we neglect the transition $3\!\rightarrow\! 2$ because $\omega_{0,1}$ are chosen to be far detuned from $\Omega_{32}$. 
If photon-$B$ is in mode $|\omega_0\rangle$, it is far off resonance from the transitions, and, using the same value of $a$ as above, $r_{11}(\omega_0)=R_{3}(\omega_0)=1$ while $r_{13}(\omega_0)=0$. 
Hence, the output state in this case is $|\phi^{\text{out}}_1(\omega_A)\rangle \otimes |\omega_B\rangle$\textemdash photon-$B$ is unaffected.
However, if photon-$B$ is in $|\omega_1\rangle$, the state after scattering is

\begin{eqnarray}
|\phi^{\text{out}}_2(\omega_A,\omega_B=\omega_1)\rangle&=&
r_{13}(\omega_A)R_{3}(\omega_B)|\tilde{\omega}_A\rangle |\omega_B\rangle|3\rangle  \nonumber \\
&+& r_{11}(\omega_A)r_{13}(\omega_B)|\omega_A\rangle |\tilde{\omega}_B\rangle|3\rangle.\;\;
\label{Phi_AB}
\end{eqnarray}

Two possible outcomes exist: (i) if the 4LS is in state
$|1\rangle$ after step $1$, photon-$B$ will be trapped, but (ii) if
the 4LS is in state $|3\rangle$, photon-$B$ is on resonance with
transition $3\!\rightarrow\!4$ and gains a $\pi$-phase [$R_{3}(\omega_1)=e^{i\pi}$]. 
The 4LS being in state $|3\rangle$ is, of course, conditioned upon photon-$A$ in step 1 being in $|\omega_1\rangle$.
Notice that the $\pi$-phase shift is independent of coupling strength $\Gamma$ which only determines the operation bandwidth of photon pulses (for details, see the discussion of fidelity below).
The robustness of the $\pi$-phase shift of the reflected photon is the result of a Fano resonance \cite{ChangNatPhy07, FanoRMP10}: the interference of paths which  bypass the 4LS with those which go through it causes the wavefunction to vanish to the right of the 4LS ($x>0$)---analogous to the well-known perfect destructive interference in transmission past a 2LS  \cite{ChangNatPhy07, FanoRMP10}---and at the same time cause the $\pi$-phase shift of the reflected photon. 

\emph{Step 3\textemdash Retrieval of A}. By sending in the output photon
from step 1, we retrieve photon-$A$. This process is the time-reversal of photon trapping. 
The full wavefunction that results is increasingly complicated; accordingly, we focus on the specific case needed---using the two conditions $2(\Omega_{12}+\Omega_{32})a=2n_1\pi$ and $2\omega_0a=(2n_0+1)\pi$---and relegate the full wavefunction, useful for other cases, to the Supplementary Material \cite{Note1}. 
The state after this step reads
\begin{eqnarray}
 |\phi^{\text{out}}_3 & (\omega_A,\omega_B)\rangle = r_{11}(\omega_A)r_{11}(\omega_B)|\omega_B\rangle\otimes
r_{11}(\omega_A)|\omega_A\rangle |1\rangle    \nonumber \\
&  +r_{11}(\omega_A)r_{13}(\omega_B)|\tilde{\omega}_B\rangle \otimes R_3(\omega_A)|\omega_A\rangle|3\rangle \nonumber \\
&  +r_{13}(\omega_A)R_3(\omega_B)|\omega_B\rangle \otimes 
r_{13}(\omega_A)|\omega_A\rangle|1\rangle \;.
\label{eq:phistep3}
\end{eqnarray}
In our case, the factors $r_{ij}$ and $R_3$ are all either $0$ or $\pm 1$ (see Table I in the Supplementary Material). The first line of Eq.\,(\ref{eq:phistep3}) corresponds to input qubits in the $00$ state, line two is for $01$, and the last line covers both $10$ and $11$.

\emph{Step 4\textemdash Retrieval of B}. In the case $\omega_A=\omega_0$ and $\omega_B=\omega_1$, photon $B$ is trapped in the 4LS in step 2. 
Time reversal arguments imply that sending in a $C$ photon of frequency $\Omega_{32}$ will release photon $B$ in this case but will simply pass through the system without interacting in the other cases. The final state after all four steps is 
\begin{eqnarray}
|\phi^{\text{out}}_4 \rangle & = &
f_1(\omega_A,\omega_B)|\omega_A\rangle|\omega_B\rangle|\omega_C\rangle|1\rangle \;
\nonumber \\
& - &   f_2(\omega_A,\omega_B)
|\omega_A\rangle|\tilde{\omega}_B\rangle |\omega_C'\rangle|1\rangle
\label{eq:phi4out}
\end{eqnarray}
where
\begin{eqnarray}
 f_1(\omega_A,\omega_B)&=&r^2_{11}(\omega_A)r_{11}(\omega_B)+r_{13}^2(\omega_A)R_3(\omega_B), \nonumber \\
 f_2(\omega_A,\omega_B)&=&r_{11}(\omega_A)R_{3}(\omega_A)r_{13}(\omega_B).
\end{eqnarray}
The second line in Eq.\,(\ref{eq:phi4out}) corresponds to an input $\omega_A=\omega_0$ and $\omega_B=\omega_1$; the state $|\omega_C'\rangle$ signifies that the frequency of the $C$ photon is now $\omega_1$---it is the retrieved $B$ photon. By filtering out the frequency $\Omega_{32}$ and relabeling $|\omega_C'\rangle$ as $|\omega_B \rangle$, we obtain the final state $|\phi_f(\omega_A \!=\! \omega_i,\omega_B \!=\! \omega_j)=(-1)^{ij}|\omega_i\rangle|\omega_j\rangle\otimes|1\rangle$, $i,j=0$ or $1$. 

Thus we see that the above steps give rise to the desired $\pi$-phase gate:
\begin{equation}
 U_{AB}=\exp\big\{i\pi|\omega_1\rangle_A\langle\omega_1| \otimes
 |\omega_1\rangle_B\langle\omega_1|\big\}.
\end{equation}
Here, we assume the use of quantum memories and direct photon $A$ from step $3$ to quantum memory $A$. Photon $B$ from either step $2$ or $4$ is directed to quantum memory $B$ after filtering out frequency $\omega_C$.

We now analyze the gate performance by considering photon pulses with a finite spectral width $\sigma$ and including atomic loss ($\Gamma^{\prime}>0$). In particular, we consider Gaussian input pulses $A$, $B$, and $C$ centered at frequencies $\omega_1$, $\omega_1$, and $\Omega_{32}$, respectively: 
\begin{eqnarray}
 |\phi_{A,B}\rangle&=&\int
d\omega_{A,B} g_\sigma(\omega_{A,B}-\omega_1) |\omega_{A,B}\rangle,\nonumber \\
|\phi_{C}\rangle&=&\int
d\omega_{C} g_\sigma(\omega_{C}-\Omega_{32}) |\omega_{C}\rangle, \quad g_\sigma(\omega)\propto e^{-\frac{\omega^2}{2\sigma^2}}.\qquad
\end{eqnarray}
The corresponding temporal width is $\Delta T=1/(2\sigma)$. After the scattering, the final state of the system is $|\phi_f\rangle=\int d\omega_Ad\omega_Bd\omega_C g_{\sigma}(\omega_A)g_{\sigma}(\omega_B)g_{\sigma}(\omega_C)|\phi^{\text{out}}_4(\omega_A,\omega_B,\omega_C)\rangle$. 
The fidelity of the photon-atom gate is given by
\begin{equation}
F \equiv |\langle \psi |\phi_f\rangle|^2,
\label{eq:F1}
\end{equation}
where 
$|\psi\rangle = -|\phi_{A}\rangle|\phi_{B}\rangle|\phi_{C}\rangle\otimes|1\rangle$ 
is the target state.

 \begin{figure}[t!]
 \centering
 \includegraphics[width=0.45\textwidth]{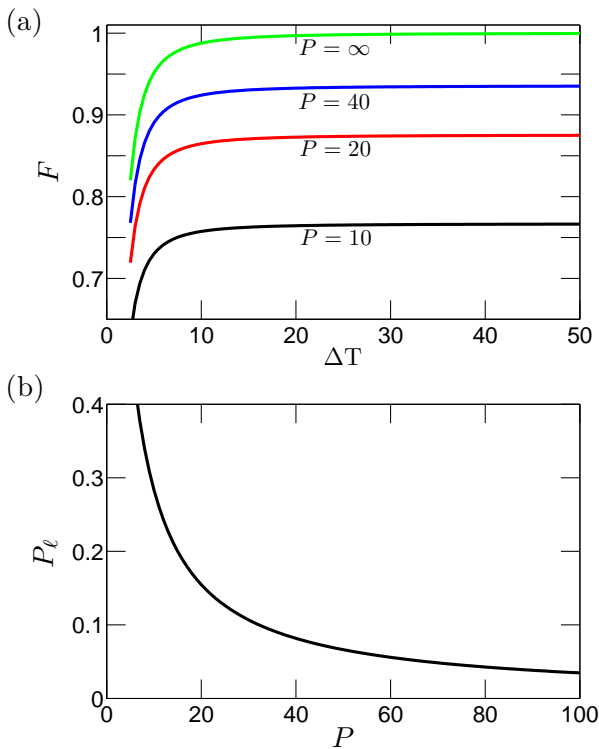}
\caption{Fidelity and leakage error of the photon-photon $\pi$-phase gate. (a) Fidelity $F$ as a function of the pulse width $\Delta
T$ (in units of $\Gamma^{-1}$) for $P=10, 20, 40, \infty$. (b) The
leakage probability $P_{\ell}$ as a function $P$ with $\Delta
T=10\Gamma^{-1}$. 
}
 \label{fig:Scheme2F}
 \end{figure}

The atomic loss is characterized by introducing the effective Purcell factor $P=\Gamma/\Gamma^{\prime}$. We note that large values of $P$ ($>20$) have been demonstrated in recent experiments using either superconducting circuits \cite{HoiPRL11}, photonic-crystal waveguides \cite{LauchtPRX12}, or semiconductor nanowires \cite{Mikkelson}. To quantify the effect of loss, we define the probability of leakage, $P_{\ell}$, to be the probability of losing the photon during the operation through spontaneous emission
\begin{equation}
P_{\ell} \equiv 1-|\langle \phi_f|\phi_f \rangle|^2.
\label{eq:Pl1}
\end{equation}

Figure \ref{fig:Scheme2F}(a) shows the fidelity of our scheme as a function of the pulse width $\Delta T$. For a short pulse, the spectral width is large, and so the fidelity is limited by the large frequency variation. As $\Delta T$ increases to $~10\Gamma^{-1}$, the fidelity starts to saturate and is only limited by the atomic loss. A fidelity of $86\%$ and $94\%$ can be achieved for $P=20$ and $P=40$, respectively. Figure \ref{fig:Scheme2F}(b) shows that the leakage error decays rapidly as $P$ increases and can be as small as a few percent for $P$ approaching $100$, which is feasible in the near future given the rapid experimental advances in 1D waveguide systems. Such a leakage error is acceptable, since it can be efficiently corrected by concatenated coding \cite{Nielsen00, DuanPRL04}.

We now make a rough estimate of the gate operation time. Since the gate fidelity is insensitive to the pulse width variation once $\Delta T$ is sufficiently large  [Fig.\,\ref{fig:Scheme2F}(a)], we choose $\Delta T \!=\! 10\Gamma^{-1}$ for practical estimation. Using a superconducting circuit as an example, we estimate the duration of our photon-photon $\pi$-phase gate  to be $~30\Gamma^{-1}\sim300\,\text{ns}$ for a superconducting qubit with $\Gamma=2\pi\!\times\! 100\,$MHz \cite{HoiPRL11}. Such an operation time is compatible with current qubit coherence times, which are on the order of $1\,\mu\text{s}$ \footnote{Recently, even higher coherence times ($10$-$20\,\mu\text{s}$ \cite{PaikPRL11} and $\sim\! 0.1\,\text{ms}$ \cite{RigettiPRB12}) have been achieved by placing the qubit in a 3D cavity.}.

An alternative $\pi$-phase gate using only a three level system (3LS) is possible by adapting the cavity-based proposal of Ref.\,\onlinecite{DuanPRL04}.
First, one constructs a $\pi$-phase gate between a photon qubit and the local qubit (3LS). 
Then, using the photon-atom $\pi$-phase gate as a building block, a $\pi$-phase gate between two photons $A$ and $B$ can be implemented by sending them into the system successively. 
This proposal has the additional advantage of naturally realizing a photon-atom $\pi$-phase gate, which can be used to entangle distant quantum nodes in a large quantum network \cite{DuanRMP10}. 
Details of the 3LS scheme can be found in the Supplementary Materials \cite{Note1}.


In summary, we demonstrate that two-qubit gates for photonic quantum computation can be designed in 1D waveguide-QED systems. 
Our waveguide-based proposal has several potential advantages over quantum computation based on cavity photons or stationary qubits. 
The operation time here is limited only by the coupling strength, while in the cavity case the cavity line width is the bottleneck. 
Also, our scheme does not require fine tuning of the interaction time, which is often a significant source of error.
In addition, our proposal is different from and has advantages over schemes based on cross-Kerr nonlinearity \cite{ShapiroPRA06}: a real transition ($1\rightarrow 3$) occurs in the four-level system rather than just transient induced polarization, and the $\pi$-phase shift is robust as the result of a Fano resonance \cite{FanoRMP10}, independent of the coupling strength and interaction time.
Overall, the system proposed here can be an important building block for future on-chip quantum networks: taking superconducting circuits as an example, we can envision such a network with (i) single photons generated using microwave resonators \cite{EichlerPRL11}, (ii) qubit photons stored in quantum memories formed from the $M$-level scheme \cite{Note1}, (iii) photon flow regulated by single-photon routers \cite{HoiPRL11}, and (iv) two-photon operations realized by our 4LS-waveguide system.

This work was supported in part by the U.S.\,Office of Naval Research.  
H.Z.\ is supported by a John T.\ Chambers Fellowship from the Fitzpatrick
Institute for Photonics at Duke University.

%

\newpage
\widetext

\begin{center}
\large\bf Supplementary Material for ``Waveguide-QED-Based Photonic Quantum Computation''
\end{center}

\global\long\def\theequation{S\arabic{equation}}

\global\long\def\thefigure{S\arabic{figure}}
\setcounter{equation}{0}
\setcounter{figure}{0}

In this Supplementary Material, we present a generalization of our scheme to polarizaton-encoded qubits, the derivation of the results in Eqs.\,(3)-(9), an extension of the scheme to a photon-qubit quantum memory, and an alternative implementation of a photon $\pi$-phase gate based on a three-level system (3LS). 

\section*{Generalization to Polarization Qubits}
The key to this generalization is to choose a four-level system (Fig.\,1 in the main text) such that both the transitions $1\rightarrow 2$ and $3\rightarrow 4$
are polarization selective, i.e., they interact with, for example, only $\pi$-polarized photons but not $\sigma$-polarized photons.
In this case, we can define our photonic polarization-qubit to be $|0\rangle=|\omega_0\rangle_{\sigma}$ and $|1\rangle=|\omega_1\rangle_{\pi}$.
Since information is stored in the polarization degree of freedom, single-qubit operations can be performed via polarization rotations.
Then all the discussion of two-qubit gates in the main text follows easily with the newly defined polarization qubits.

\section*{Derivation of the Output States in Eqs.\,(3)-(9)}
For an incoming left-going photon of frequency $\omega_A$ and initialized 4LS in state $|1\rangle$, the single-photon scattering eigenstate can be written as
\begin{eqnarray}
\label{eq:psi1}
&|\psi_1\rangle=\int dx[\phi_{1R}(x)a^{\dagger}_R(x)+\phi_{1L}(x)a^{\dagger}_L(x)]|1,\emptyset\rangle+e_2|2,\emptyset\rangle \nonumber \!\!\\
&+\int dx[\phi_{3R}(x)a^{\dagger}_R(x)+\phi_{3L}(x)a^{\dagger}_L(x)]|3,\emptyset\rangle, 
\end{eqnarray}
where $|i,\emptyset\rangle$ is the vacuum state of photons with the 4LS in state $|i\rangle$.
The Schr\"{o}dinger equation $H|\psi_1\rangle=\hbar\omega_A|\psi_1\rangle$ gives
\begin{subequations}
\label{eq:SchEq}
 \begin{eqnarray}
&&\!\!\!\! \Big(-i\frac{d}{dx}-\omega_A\Big)\phi_{1R}(x)+V\delta(x)e_2=0,\\
&&\!\!\!\!  \Big(i\frac{d}{dx}-\omega_A\Big)\phi_{1L}(x)+V\delta(x)e_2=0,\\
&&\!\!\!\!  \Big(-i\frac{d}{dx}+\Omega_3-\omega_A\Big)\phi_{3R}(x)+V\delta(x)e_2=0,\\
&&\!\!\!\!  \Big(i\frac{d}{dx}+\Omega_3-\omega_A\Big)\phi_{3L}(x)+V\delta(x)e_2=0,\\
&&\!\!\!\! \Big(\Omega_2-\frac{i\Gamma^{\prime}}{2}-\omega_A\Big)e_2+V\sum_{i=1,3}[\phi_{iR}(0)+\phi_{iL}(0)]=0.\quad\quad\;\;
 \end{eqnarray}
\end{subequations}
We assume the following solution ansatz
\begin{subequations}
\label{eq:Ansatz}
 \begin{eqnarray}
  \phi_{1R}(x)&=&e^{ik_1x}[\theta(-x)+\beta_{1R}\theta(x)],\\
  \phi_{1L}(x)&=&e^{-ik_1x}[\alpha_{1L}\theta(-x)+\beta_{1L}\theta(x)],\\
   \phi_{3R}(x)&=&e^{ik_3x}\beta_{3R}\theta(x),\\
  \phi_{3L}(x)&=&e^{-ik_3x}[\alpha_{3L}\theta(-x)+\beta_{3L}\theta(x)],
 \end{eqnarray}
\end{subequations}
where $k_1=\omega_A$ and $k_3=\omega_A-\Omega_3$ (we set $c=1$). Substituting Eq.\,(\ref{eq:Ansatz}) into Eq.\,(\ref{eq:SchEq}), setting $\phi_{iR/L}(0)=[\phi_{iR/L}(0^{+})+\phi_{iR/L}(0^{-})]/2$, and imposing the hard-wall boundary condition at the end of the waveguide
\begin{equation}
 \phi_{1R}(a)+\phi_{1L}(a)=\phi_{3R}(a)+\phi_{3L}(a)=0,
\end{equation}
we obtain the following solution
\begin{eqnarray}
\label{eq:Sol1}
\alpha_{1L}&=&e^{2i\omega_A
a}\frac{-\Omega_{2}+\frac{i\Gamma^{\prime}}{2}+\omega_A-\frac{i\Gamma}{2}[e^{2i\tilde{\omega}_Aa}-e^{-2i\omega_A
a}]}{\Omega_{2}-\frac{i\Gamma^{\prime}}{2}-\omega_A+\frac{i\Gamma}{2}[e^{2i\tilde{\omega}_Aa}+e^{2i\omega_A
a}-2]}, \nonumber \\
\alpha_{3L}&=&\frac{(i\Gamma/2)(e^{2i\omega_A
a}-1)[e^{2i\tilde{\omega}_Aa}-1]}{\Omega_{2}-\frac{i\Gamma^{\prime}}{2}-\omega_A+\frac{i\Gamma}{2}[e^{2i\tilde{\omega}_Aa}+e^{2i\omega_A
a}-2]} \;,
\end{eqnarray}
where $\tilde{\omega}_A=\omega_A-\Omega_{3}$. According to the Lippmann-Schwinger formalism \cite{SakuraiQM94, ShenPRA07, ZhengPRA10-2}, we read off the ``out'' state after scattering from the scattering eigenstate in Eq.\,(\ref{eq:psi1}). 
The resulting asymptotic output state $|\phi^{\text{out}}_1(\omega_A)\rangle$ of photon $A$ is then given in Eq.\,(2) in the main text with $r_{11}$ and $r_{13}$ corresponding to $\alpha_{1L}$ and $\alpha_{3L}$, respectively.

As noted in the main text, for the operation of our gate, we use the two conditions $2(\Omega_{12}+\Omega_{32})a=2n_1\pi$ and $2\omega_0a=(2n_0+1)\pi$. The first assures complete transfer of the 4LS from state 1 to state 3 when a photon at $\omega_1$ is incident; the second less critical criterion ensures that certain phase factors are real. In following the photon sequences, it is useful to have in mind the reflection coefficients evaluated for these conditions. See Table I for the values used in our photon gate scheme.

\begin{table}[t]
\begin{tabular}{|c|c|c|c|}
 \quad \quad \quad & \quad $\omega_0$ \quad & \quad $\omega_1$ \quad & \quad $\Omega_{32}$ \quad \\
 \hline
$r_{11}$ & 1 & 0 & - \\
$r_{13}$ & 0 & $\!\!\!-1$ & 0 \\
$r_{31}$ & 0 & - & $\!\!\!-1$ \\
$r_{33}$ & 1 & - & 0 \\
$R_3$ & 1 & $\!\!\!-1$ & - \\
\end{tabular}
\caption{Table of values of $r_{11}$, $r_{13}$, $r_{31}$, $r_{33}$, and $R_3$ at $\omega_0$, $\omega_1$, and $\Omega_{32}$ when the two conditions $2(\Omega_{12}+\Omega_{32})a=2n_1\pi$ and $2\omega_0a=(2n_0+1)\pi$ hold. Only those used in our scheme are shown.
}
\end{table}

Next, in Step 2 we send in the second photon $B$ of frequency $\omega_B$. 
If the 4LS is in state $|1\rangle$, then photon-$B$ will scatter in the same way as the first photon and the output state is $|\phi^{\text{out}}_1(\omega_B)\rangle$.
If the 4LS is in state $|3\rangle$ and we neglect the transition $3\rightarrow2$ since photon-$B$ is far off resonance from it, then photon-$B$ will interact with the transition $3\rightarrow4$ only. 
The single-photon scattering eigenstate takes the form 
\begin{eqnarray}
 \label{eq:phi2}
|\psi_2\rangle&=&\int dx[\phi_{R}(x)a^{\dagger}_R(x)+\phi_L(x)a^{\dagger}_L(x)]|3,\emptyset\rangle \nonumber \\
 && \qquad+e_4|4,\emptyset\rangle. 
\end{eqnarray}
From the Sch\"{o}dinger equation $H|\psi_2\rangle=\hbar\omega_B|\psi_2\rangle$, we have
\begin{subequations}
\begin{eqnarray}
 \label{eq:SchEq2}
 &&\Big(-i\frac{d}{dx}-\omega_B \Big)\phi_R(x)+V\delta(x)e_4=0,\\
 &&\Big(i\frac{d}{dx}-\omega_B \Big)\phi_L(x)+V\delta(x)e_4=0, \\
 &&\Big(\Omega_{34}-\frac{i\Gamma^{\prime}}{2}-\omega_B \Big)e_4+V[\phi_R(0)+\phi_L(0)]=0.\quad
\end{eqnarray}
\end{subequations}
Again, we assume the following ansatz
\begin{subequations}
\begin{eqnarray}
 \label{eq:ansatz2}
 &&\phi_R(x)=e^{ik_4x}[\theta(-x)+\beta_R\theta(x)],\\
 &&\phi_L(x)=e^{-ik_4x}[\alpha_L\theta(-x)+\beta_L\theta(x)],
\end{eqnarray}
\end{subequations}
where $k_4=\omega_B$. Using the hard-wall boundary condition $\phi_R(a)+\phi_L(a)=0$, it is straightforward to obtain the following solution
\begin{eqnarray}
 \label{eq:Sol2}
 \beta_R&=&-e^{-2i\omega_Ba}\beta_L=\frac{\Omega_{34}-\frac{i\Gamma^{\prime}}{2}-\omega_B}{\Omega_{34}-\frac{i\Gamma^{\prime}}{2}-\omega_B-\frac{i\Gamma}{2}(1-e^{2i\omega_Ba})},\nonumber \\
 \alpha_L&=&\frac{-(\Omega_{34}-\frac{i\Gamma^{\prime}}{2}-\omega_B)e^{2i\omega_B a}+\frac{i\Gamma}{2}(1-e^{2i\omega_B a})}{\Omega_{34}-\frac{i\Gamma^{\prime}}{2}-\omega_B-\frac{i\Gamma}{2}(1-e^{2i\omega_B a})}.
\end{eqnarray}
Again, we read off the output state of photon-$B$ for the case 4LS in state $|3\rangle$ as $R_3(\omega_B)|\omega_B\rangle\otimes|3\rangle$ with $R_3$ corresponding to $\alpha_L$.
Combining the output states from both cases 4LS in states $|1\rangle$ and $|3\rangle$, we obtain the general output state after Step 2 in Eq.\,(4) in the main text.

Step 3 consists of sending in the output $A^{\prime}$ photon from Step 1 in order to retrieve photon-$A$. By using the same approach as for Steps 1 and 2 above, one finds that the state after this step reads
\begin{eqnarray}
 |\phi^{\text{out}}_3(\omega_A,\omega_B)\rangle \!&=\!&r_{11}(\omega_A)r_{11}(\omega_B)|\omega_B\rangle\otimes|\phi^{\text{out}}_1(\omega_A)\rangle \nonumber \\
&+&r_{11}(\omega_A)r_{13}(\omega_B)|\tilde{\omega}_B\rangle \otimes [R_3(\omega_A)|\omega_A\rangle|3\rangle] \nonumber \\
&+&r_{13}(\omega_A)R_3(\omega_B)|\omega_B\rangle \otimes |\phi_3(\tilde{\omega}_A)\rangle,
\end{eqnarray}
where $|\phi^{\text{out}}_1(\omega_A)\rangle$ is given in Eq.\,(2) in the main text and
\begin{equation}
 |\phi_3(\omega)\rangle=r_{33}(\omega)|\omega\rangle\otimes|3\rangle+r_{31}(\omega)|\bar{\omega}\rangle\otimes|1\rangle,
\end{equation}
with $\bar{\omega}=\omega+\Omega_{13}$,
\begin{eqnarray}
&&r_{33}(\omega)=e^{2i\omega
a}\frac{-\Omega_{32}+\frac{i\Gamma^{\prime}}{2}+\omega-\frac{i\Gamma}{2}[e^{2i\bar{\omega}a}-e^{-2i\omega
a}]}{\Omega_{32}-\frac{i\Gamma^{\prime}}{2}-\omega+\frac{i\Gamma}{2}[e^{2i\bar{\omega}a}+e^{2i\omega
a}-2]}, \nonumber \\
&&r_{31}(\omega)=\frac{(i\Gamma/2)(e^{2i\omega
a}-1)[e^{2i\bar{\omega}a}-1]}{\Omega_{32}-\frac{i\Gamma^{\prime}}{2}-\omega+\frac{i\Gamma}{2}[e^{2i\bar{\omega}a}+e^{2i\omega
a}-2]} .
\end{eqnarray}
When the two conditions $2(\Omega_{12}+\Omega_{32})a=2n_1\pi$ and $2\omega_0a=(2n_0+1)\pi$ introduced in Step 1 hold, these reflection amplitudes can be either $0$, $+1$, or $-1$, as indicated in Table I. 

Finally, in Step 4, we send in an auxiliary photon-$C$ with frequency $\omega_C=\Omega_{32}$. This allows the recovery of photon $B$, which in the case $\omega_A=\omega_0$ and $\omega_B=\omega_1$ is trapped in the 4LS in Step 2. 
According to time reversal argument, sending in a $C$ photon of frequency $\Omega_{32}$ will release photon $B$. A very similar analysis shows that the final state after all four steps is
\begin{eqnarray}
\lefteqn{ |\phi^{\text{out}}_4(\omega_A,\omega_B,\omega_C)\rangle = f_1(\omega_A,\omega_B)|\omega_A\rangle|\omega_B\rangle|\omega_C\rangle|1\rangle } \qquad\qquad\qquad\qquad\qquad\qquad\qquad\qquad\qquad\qquad\qquad \nonumber \\
\lefteqn{ \qquad\qquad\qquad\qquad + [ f_2(\omega_A,\omega_B)|\omega_A\rangle|\tilde{\omega}_B\rangle} \qquad\qquad\qquad\qquad\qquad\qquad\qquad\qquad\qquad\qquad\qquad  \nonumber \\
\lefteqn{ \qquad\qquad\qquad +f_3(\omega_A,\omega_B)|\tilde{\omega}_A\rangle|\omega_B\rangle ]|\phi_3(\omega_C)\rangle.} \qquad\qquad\qquad\qquad\qquad\qquad\qquad\qquad\qquad\qquad\qquad
\label{eq:phi4out}
\end{eqnarray}
where
\begin{eqnarray}
 f_1(\omega_A,\omega_B)&=&r^2_{11}(\omega_A)r_{11}(\omega_B)+r_{13}(\omega_A)r_{31}(\tilde{\omega}_A)R_3(\omega_B)], \nonumber \\
f_2(\omega_A,\omega_B)&=&r_{11}(\omega_A)R_{3}(\omega_A)r_{13}(\omega_B) \nonumber \\ 
f_3(\omega_A,\omega_B)&=&r_{11}(\omega_A)r_{13}(\omega_A)r_{11}(\omega_B) \nonumber \\
   &+& r_{13}(\omega_A)r_{33}(\tilde{\omega}_A)R_{3}(\omega_B).
\end{eqnarray}
Under our two conditions, these equations are equivalent to that given in the main text (see Table I for the values of the reflection amplitudes), thus establishing the photon phase gate, Eq.\,(10) of the main text.

\section*{Photon-Qubit Quantum Memory}
\begin{figure}[t]
 \centering
 \includegraphics[width=0.30\textwidth]{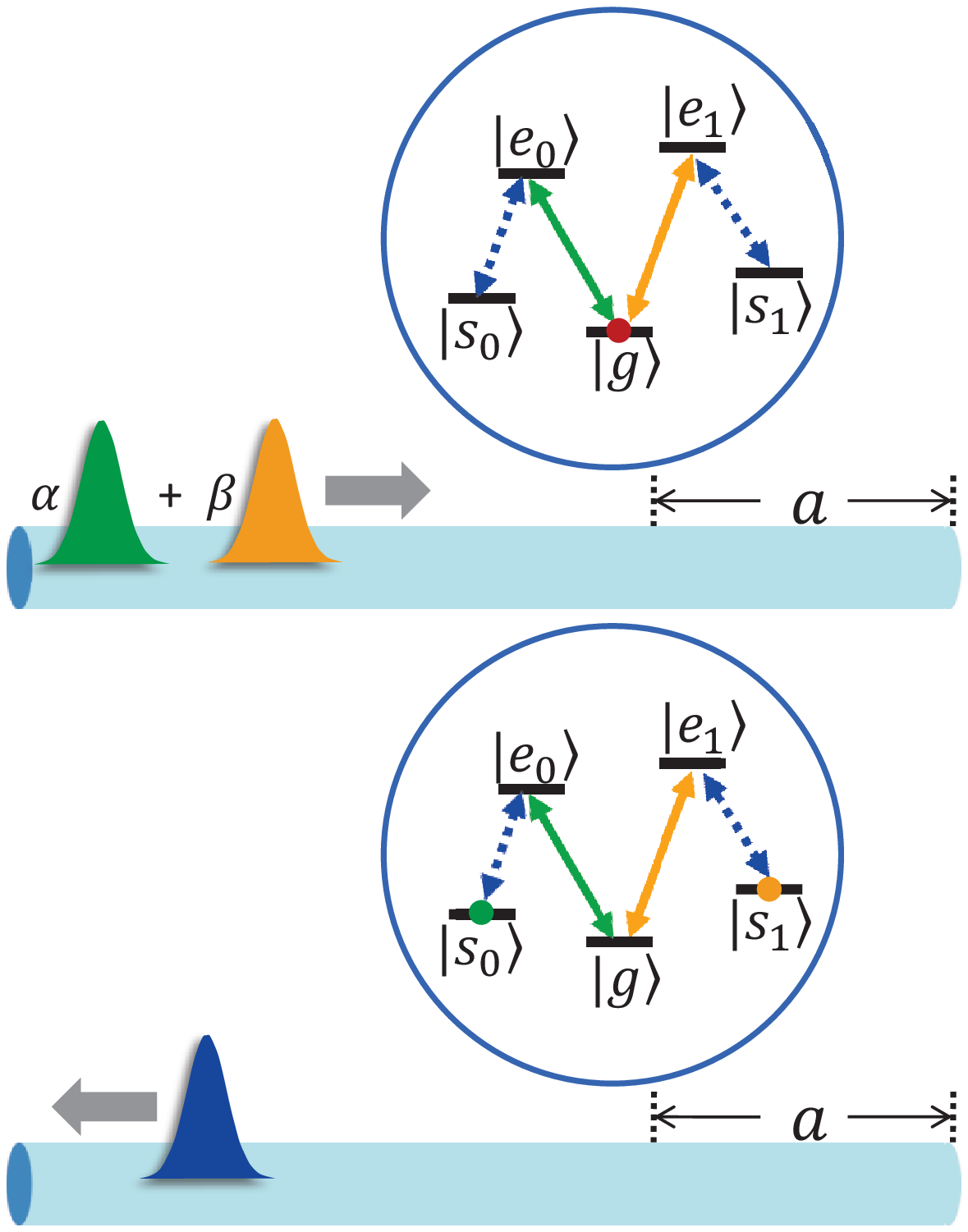}
\caption{Schematic diagram of an $M$-type five-level system as a quantum memory of qubit photons. 
          The top and bottom panels show the initial and final states of mapping a qubit-photon state to a matter-qubit state.
          An arbitrary superposition of $0$-photon (green pulse) and $1$-photon (yellow pulse) is store in the matter-qubit states $|s_0\rangle$ and $|s_1\rangle$. 
          Meanwhile, an anxiliary photon-$C$ (blue pulse) is emitted after the mapping.
         }
 \label{fig:M_5LS}
 \end{figure}

We extend the 4LS scheme to an M-type five-level system (5LS) to construct a quantum memory of the qubit photons, as shown in Fig.\,\ref{fig:M_5LS}.
The 5LS is chosen so that the transition energies of the ground state $|s\rangle$ to the two excited states $|e_0\rangle$ and $e_1\rangle$ match the photon-qubit frequencies $\omega_0$ and $\omega_1$, respectively. 
Because $\omega_0$ and $\omega_1$ are well separated in frequency ($|\omega_0-\omega_1|\gg\Gamma$), the $|\omega_0\rangle$ ($|\omega_1\rangle$) photon only interacts with the $g\rightarrow e_0$ ($g\rightarrow e_1$) transition. 
In addition, the $s_0\rightarrow e_0$ and $s_1\rightarrow e_1$ transitions have the transition energy $\omega_{es}$. 
According to our calculations in the trapping step in the main text, the output state of an incoming $|\omega_0\rangle$ photon with the 5LS initialized in ground state $|g\rangle$ reads 
\begin{equation}
|\phi_0\rangle=|\omega_{es}\rangle \otimes|s_0\rangle.
\end{equation}
Similarly, the output state of an incoming $|\omega_1\rangle$ photon is
\begin{equation}
|\phi_1\rangle=|\omega_{es}\rangle \otimes|s_1\rangle.
\end{equation}
Hence, the output state after scattering an arbitrary photon-qubit state $\alpha|\omega_0\rangle+\beta|\omega_1\rangle$ reads
\begin{equation}
|\phi_{\text{out}}\rangle=(\alpha|s_0\rangle+\beta|s_1\rangle)\otimes|\omega_{es}\rangle.
\end{equation}
Therefore, the photon-qubit  is mapped into a matter-qubit made of two metastable states $|s_0\rangle$ and $|s_1\rangle$; at the same time, an anxiliary photon-$C$ of frequency $\omega_{es}$ is emitted. 
Retrieval of the photon-qubit can be done simply by sending in the anxiliary photon-$C$.
 
Similarly, the generalization of this quantum memory to polarization-encoded photon qubits is straightforward: we choose the 5LS such that the transitions $g\rightarrow e_0$ and $g\rightarrow e_1$ only interact with $\pi$-polarized and $\sigma$-polarized photons respectively. 
Defining the $\sigma$-polarized ($\pi$-polarized) photons as the $0$-photon ($1$-photon) qubit states, the above scheme will work as a quantum memory of the polarization-encoded photonic qubits.

\section*{Alternative 3LS Scheme}
\begin{figure}[t]
 \centering
 \includegraphics[width=0.46\textwidth]{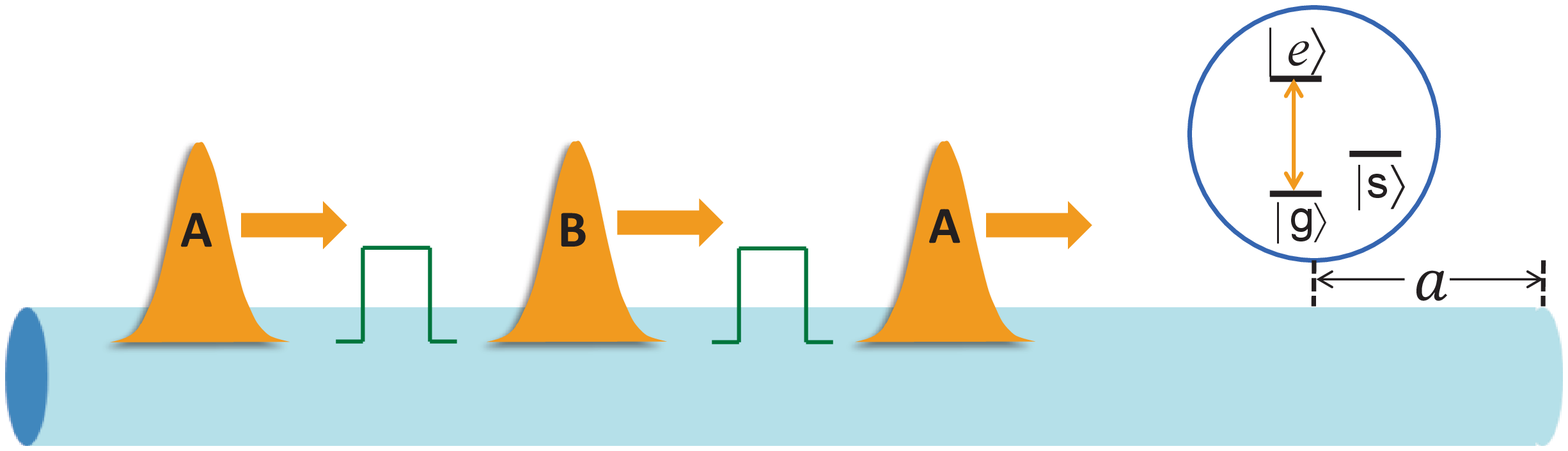}
\caption{Schematic diagram of the operation sequence of the
$\pi$-phase gate between two photons $A$ and $B$. The three-level system (3LS) is located a distance $a$ from the end of the semi-infinite waveguide and is initialized in an equal superposition of the $g$ and $s$ states. $A$ and $B$ are reflected successively from the semi-infinite waveguide coupled to the 3LS. Between the reflections, single-qubit rotation pulses (green rectangles) are applied to the atomic qubit made of states $|g\rangle$ and
$|s\rangle$.}
 \label{fig:SchemeA}
 \end{figure}

In this realization (Fig.\,\ref{fig:SchemeA}), we consider a semi-infinite 1D waveguide side-coupled to a 3LS, which is located a distance $a$ from the end. Such a system could be realized in a variety of experimental systems \cite{AbdumalikovPRL10-2,BianchettiPRL10,EichlerPRL11-2,HoiPRL12-2}. 
A $\pi$-phase gate between two photons $A$ and $B$ is realized by reflecting them from the end of the waveguide.

The Hamiltonian of the system is given by
\begin{eqnarray}
H_1&=&H_{\rm wg}+H_{ee}+\!\!\!\sum_{\alpha=R,L}\int\!\!
dx\hbar V \delta(x)[a_{\alpha}^{\dagger}(x)\sigma_{ge}+\textrm{h.c.}], \nonumber \\
H_{\rm wg}&=&\int dx\frac{\hbar
c}{i}\Big[a_{R}^{\dagger}(x)\frac{d}{dx}a_{R}(x)-a_{L}^{\dagger}(x)\frac{d}{dx}a_{L}(x)\Big],
\end{eqnarray}
where $H_{ee}=\hbar(\Omega_{eg}-i\Gamma^{\prime}/2)\sigma_{ee}$ and
$\sigma_{ij}\equiv |i\rangle\langle j|$. The transition from the ground state $|g\rangle$ to the excited state $|e\rangle$ couples to the
waveguide modes ($a_{R,L}$); an imaginary term models the loss of the excited state at rate $\Gamma^{\prime}$. The decay rate to the waveguide continuum is $\Gamma=2V^2/c$, where $c$ is the group velocity of photons. For simplicity, we set $\hbar=c=1$. Note that the metastable state $|s\rangle$ does not appear in the Hamiltonian as it is decoupled from the waveguide; however, its presence is essential since $|g\rangle$ and $|s\rangle$ form the atomic qubit. As in the main text, we consider a photonic qubit coded in the frequency domain.

Our first step is to realize a photon-atom $\pi$-phase gate. 
Consider a single incoming photon with frequency $\omega$. In the
case that the three-level system is in $|g\rangle$, we obtain the
output state by imposing a hard-wall boundary condition at the end
of the waveguide, thus giving
\begin{eqnarray}
|\phi^{\text{out}}_g(\omega)\rangle&=&r_g(\omega)|\omega\rangle_L,
\nonumber \\
r_g(\omega)&=&\frac{-(\Delta-\frac{i\Gamma^{\prime}}{2})e^{2i\omega
a}+\frac{i\Gamma}{2}(1-e^{2i\omega
a})}{\Delta-\frac{i\Gamma^{\prime}}{2}-\frac{i\Gamma}{2}(1-e^{2i\omega
a})},\qquad
\end{eqnarray}
where $\Delta=\Omega_{eg}-\omega$ is the detuning. We first
illustrate the operation principle for the lossless case
$\Gamma^{\prime}=0$ and then later analyze the effect of loss in detail. 
In the lossless case, we always have perfect
reflection\textemdash$|r_g(\omega)|^2=1$\textemdash because the waveguide
is semi-infinite. Choosing two frequencies
$\omega_1=\Omega_{eg}$ and $|\omega_0-\Omega_{eg}|\gg \Gamma$, we
have $r_g(\omega_1)=-1$ and $r_g(\omega_0)=-e^{2i\omega_{0}a}$. On the
other hand, if the qubit is in $|s\rangle$, the photon gains a
trivial phase shift $r_s(\omega)=-e^{2i\omega a}$. Therefore, under
the conditions $2\omega_0a=(2n_0+1)\pi$ and
$2\Omega_{eg}a=(2n_1+1)\pi$ with $n_0,n_1\in \mathbb{Z}$, we realize
a $\pi$-phase gate between the photonic qubit ($|\omega_0\rangle$,
$|\omega_1\rangle$) and the atomic qubit ($|g\rangle$, $|s\rangle$):
\begin{eqnarray}
r_g(\omega_0)=r_s(\omega_0)=r_s(\omega_1)=-r_g(\omega_1)=1,
\nonumber \\
U_{\text{photon-atom}}=\exp\big\{i\pi
|\omega_1\rangle\langle\omega_1| \otimes |g\rangle\langle g|\big\}.
\end{eqnarray}

Using the photon-atom $\pi$-phase gate as a building block, we can implement a
$\pi$-phase gate between two photons $A$ and $B$ as in the cavity-based proposal of Ref.\,\onlinecite{DuanPRL04-2}. First, initialize the atom in the state
$|\phi_a\rangle=(|g\rangle+|s\rangle)/\sqrt{2}$. Next, send in
photon-$A$ followed by a $\pi/2$ rotation on the atom. Third, send
in photon-$B$ followed by a $-\pi/2$ rotation on the atom. Finally,
send in photon-$A$ again. This procedure produces a $\pi$-phase gate
\begin{equation}
   U_{AB}=\exp \big\{i\pi|\omega_1\rangle_A\langle\omega_1|\otimes
|\omega_1\rangle_B\langle\omega_1| \big\} \;.
\end{equation}
Our scheme closely resembles
the cavity-based proposal \cite{DuanPRL04-2}, but we rely on a
different mechanism to generate the $\pi$-phase shift in a
cavity-free setting. As for the 4LS in the main text, this phase gate requires fine tuning so that, as noted above, $2\Omega_{eg}a=(2n_1+1)\pi$; this is possible using superconducting qubits, for instance, for which the transition frequencies can be easily tuned using external magnetic flux \cite{ClarkeNat08}.

\begin{figure}[t]
 \centering
 \includegraphics[width=0.45\textwidth]{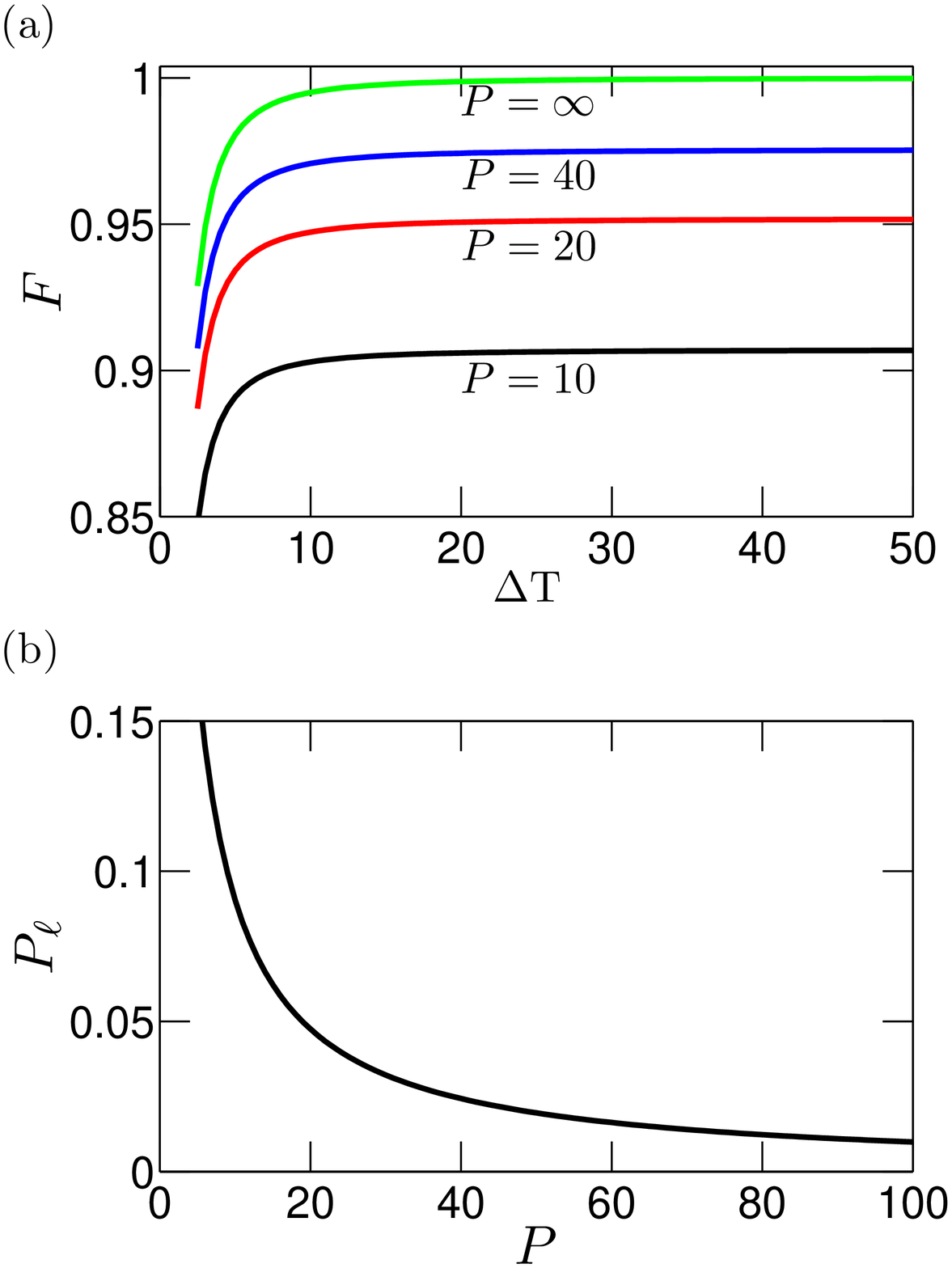}
\caption{Fidelity and leakage error of the photon-atom gate in
scheme 1. (a) Fidelity $F$ as a function of the pulse width $\Delta
T$ for four different cases. (b) The leakage probability $P_{\ell}$ as a function of
the effective Purcell factor $P$ with a pulse width $\Delta
T=10\Gamma^{-1}$.
}
\label{fig:Scheme1F}
\end{figure}

To analyze the gate performance in this scheme, we consider photon pulses with a
finite spectral width $\sigma$ and include atomic loss
($\Gamma^{\prime}>0$). In particular, we consider a Gaussian input
pulse centered at frequency $\omega_1$: 
\begin{eqnarray}
   |\phi_{i}\rangle & = &\int d\omega g_\sigma(\omega)|\omega\rangle \otimes |\phi_a\rangle \;, \\
 & & 
g_\sigma(\omega)\propto \exp\{-(\omega-\omega_1)^2/2\sigma^2\}.
\end{eqnarray}
The temporal width is $\Delta T=1/(2\sigma)$. After the scattering,
the final state of the system is 
\begin{eqnarray}
   |\phi_f\rangle & = & \int d\omega g_{\sigma}(\omega) \\
& & \times \big\{ r_g(\omega)|\omega\rangle \otimes |g\rangle +
r_s(\omega)|\omega\rangle \otimes |s\rangle \big\}/\sqrt{2} .
\nonumber
\end{eqnarray}
The
fidelity of the photon-atom gate is given by
\begin{equation}
F \equiv |\langle \psi |\phi_f\rangle|^2=\Big|\frac{1}{2}\int d\omega
g^2_{\sigma}(\omega) [r_g(\omega)-r_s(\omega)]\Big|^2, \label{eq:F1}
\end{equation}
where 
$|\psi\rangle = |\omega_1\rangle \otimes (-|g\rangle+|s\rangle)/\sqrt{2}$ 
is the target state. The atomic loss is characterized by introducing the effective Purcell factor $P=\Gamma/\Gamma^{\prime}$. To measure the effect of loss quantitatively, we define the probability of leakage, $P_{\ell}$, as the probability of losing the photon during the operation through spontaneous emission:
\begin{equation}
P_{\ell} \equiv 1-|\langle \phi_f|\phi_f \rangle|^2=1-\Big|\int d\omega
g^2_{\sigma}(\omega)\frac{[|r_g(\omega)|^2+1]}{2}\Big|^2.\label{eq:Pl1}
\end{equation}

Figure \ref{fig:Scheme1F}(a) shows the fidelity $F$ as a function of pulse temporal width $\Delta T$. For a short pulse, the spectral width is large, and so the fidelity is limited by the large frequency variation of the conditional phase $r_g(\omega)$. As $\Delta T$ increases to $~10\Gamma^{-1}$, the fidelity starts to saturate and is only limited by the atomic loss. A high fidelity ($\ge95\%$) can be achieved for a practical value of $P\ge20$. Figure \ref{fig:Scheme1F}(b) shows that the leakage probability decreases quickly as one increases $P$ and is on the order of a few percent for $P\ge20$.  Further improvement in both fidelity and loss can be expected from the rapidly development of 1D waveguide technology and schemes using off-resonance mechanisms to reduce the loss.

\end{document}